\documentclass[10pt,journal,letterpaper,twocolumn]{IEEEtran}
\usepackage[american]{babel}
\addto\captionsamerican{}
\usepackage[T1]{fontenc}
\usepackage{microtype}
\usepackage{amsmath,amssymb,mathtools}
\usepackage{graphicx,xurl}
\usepackage{cite}
\usepackage[hidelinks,hypertexnames=false]{hyperref}
\AddToHook{env/figure/begin}{}
\newtheorem{thm}{Theorem}
\newtheorem{lem}[thm]{Lemma}
\newtheorem{cor}[thm]{Corollary}
\usepackage[capitalise,noabbrev,nameinlink]{cleveref}
\crefname{figure}{Fig.}{Figs.}
\Crefname{figure}{Fig.}{Figs.}
\crefname{thm}{Theorem}{Theorems}
\Crefname{thm}{Theorem}{Theorems}
\crefname{lem}{Lemma}{Lemmas}
\Crefname{lem}{Lemma}{Lemmas}
\crefname{cor}{Corollary}{Corollaries}
\Crefname{cor}{Corollary}{Corollaries}
\AddToHook{cmd/@begintheorem/after}{\MakeLinkTarget{thm}}
\AddToHook{cmd/@opargbegintheorem/after}{\MakeLinkTarget{thm}}
\newcommand{\qed}{}
\newenvironment{pf}{\begin{IEEEproof}}{\end{IEEEproof}}
\date{}
\hypersetup{pdftitle={Dissipation Bounds and Hardness in Globally Asymptotically Stable Quadratic Systems},pdfauthor={Subhransu S. Bhattacharjee}}
\title{Dissipation Bounds and Hardness in\\Globally Asymptotically Stable Quadratic Systems}
\author{Subhransu S. Bhattacharjee\\
\normalsize The Australian National University, Canberra ACT 2600, Australia\thanks{Correspondence: \href{mailto:Subhransu.Bhattacharjee@anu.edu.au}{\nolinkurl{Subhransu.Bhattacharjee@anu.edu.au}}.}}
\begin{document}
\maketitle
\begin{abstract}
Approximating the maximum retained energy fraction in globally asymptotically stable quadratic systems of growing dimension is NP-hard at every fixed positive additive error below $1/2$, even when strict quartic Lyapunov functions and strictness proofs are constructed in polynomial time and included in the input. Polynomial-factor approximation is also NP-hard for the least dissipated fraction, energy-halving time and structured coefficient distance to loss of asymptotic stability. For a partially damped family, minimum unit-sphere coupling norm equals that distance, with two-sided inverse-square halving-time bounds at distances in $(0,1/2]$ and explicit coefficient factors. Additive hardness persists under structured coefficient perturbations and direct damping, even with global exponential stability, a strict quadratic Lyapunov function and a Hurwitz Jacobian.
\end{abstract}

\section{Introduction}
Reduced fluid and plasma models combine quadratic energy redistribution with linear damping \cite{kaptanoglu2021promoting}. Global asymptotic stability (GAS) guarantees convergence, but Lyapunov bounds may overestimate energy-halving times.

For Euclidean energy $E(z)=\lVert z\rVert^2/2$, $M_T$ denotes the maximum of $E(z(T))/E(z(0))$ at $T\geq0$ over unit full-state initial conditions, including arbitrary damped coordinates. Thus $1-M_T$ is the least dissipated fraction; $M_T\leq1/2$ guarantees halving. We ask whether strict Lyapunov certificates permit accurate polynomial-time estimation.

\subsection{Related work}
Convex bounds \cite{fantuzzi2020bounding} cover $M_T$, using observable $\lVert z\rVert^2$ and the unit initial sphere (\Cref{eq:results-detail-9}). Convergent moment and sum-of-squares (SOS) hierarchies bound minimum perturbation intensity for unsafe reachability \cite{miller2026crash}.

We adapt classical strict Lyapunov constructions \cite{malisoff2009constructions,mazenc2004strong}. Goyal et al.\ \cite{goyal2025guaranteed} characterize strict Euclidean energy decay in quadratic models with energy-conserving nonlinear terms, including our damped family (\Cref{prop:damping}). Ahmadi's Boolean quartic encoding gives cubic stability hardness \cite{ahmadi2012difficulty}; degree two was open in 2013~\cite{ahmadi2013complexity}. Quadratization reproduces trajectories on an invariant graph \cite{cai2024dissipative}; stable added directions permit center-manifold transfer of local stability \cite{carr1981applications}. Those trajectory correspondences alone do not bound full-state Euclidean energy for arbitrary auxiliary initial states. Computing structured singular values is NP-hard \cite{blondel2000survey,braatz1994complexity,fu1997hardly}; linear stability radii quantify coefficient uncertainty \cite{hinrichsen1986structured}, and overdetermined polynomial condition numbers relate residuals and coefficient distances \cite{burgisser2019homology}.

\begin{figure}[t]
\centering

\includegraphics[width=\columnwidth]{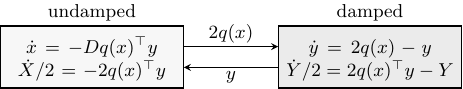}
\caption{Energy exchange in system~\labelcref{eq:field}. With $X=\lVert x\rVert^2$ and $Y=\lVert y\rVert^2$, the exchanged power $2q(x)^\top y$ cancels, giving $\dot E=-Y$ (\Cref{lem:energy}).}
\label{fig:exchange}
\end{figure}

\subsection{Contributions}
We construct quadratic systems that are GAS on every input, yet retained-energy estimation is NP-hard at every fixed positive additive error below $1/2$, even with a strict quartic Lyapunov function and rational SOS proof in the input (\Cref{thm:allstable}). The objective ranges over the full Euclidean unit sphere, including arbitrary auxiliary initial states.

With partial damping, minimum unit-sphere coupling norm equals the structured coefficient distance to loss of GAS and gives inverse-square halving-time bounds for distances in $(0,1/2]$ (\Cref{prop:distance,thm:law}). For this family, LAS and GAS verification are $\forall\mathbb R$-complete; for GAS members, degree four is minimal for strict global Lyapunov polynomials (\Cref{prop:degree,thm:real}).

The performance gap persists in the coefficient ball of \Cref{cor:robust}. Direct damping retains additive hardness with global exponential stability, a strict quadratic Lyapunov function and a Hurwitz Jacobian (\Cref{prop:damping}). These quadratic examples cover every fixed polynomial degree bound at least two (\Cref{lem:extension}).

\subsection{Preliminaries and notation}
Norms are Euclidean; $\lVert G\rVert_F^2=\sum_{j,k}G_{jk}^2$. Symbols $D$ and $\langle\cdot,\cdot\rangle_F$ denote the Jacobian and Frobenius inner product. Hurwitz matrices have negative eigenvalue real parts. Local asymptotic stability (LAS) requires Lyapunov stability and local convergence to zero; GAS adds global convergence and forward completeness (existence for all future times) \cite{haddad2008nonlinear}. For $\dot z=F(z)$, a strict global Lyapunov polynomial $V$ is radially unbounded, with $V(0)=0$, $V(z)>0$ and $\dot V(z)=\nabla V(z)^\top F(z)<0$ for $z\ne0$. A non-strict Lyapunov polynomial satisfies the same conditions with $\dot V\leq0$ \cite{malisoff2009constructions}. An SOS polynomial is a sum of squared polynomials; rational weighted SOS uses rational polynomials and non-negative rational weights.

Rational data, time horizons and variable indices are binary encoded; variable-count conventions appear below. Dimensions grow with description length; running time counts bit operations. P consists of problems decidable in deterministic polynomial time. NP consists of decision problems solvable in non-deterministic polynomial time, equivalently those with polynomial-length witnesses verifiable in polynomial time; coNP consists of their complements. A polynomial-time many-one reduction preserves answers; every NP problem reduces to an NP-hard target \cite{blondel2000survey}. Estimation reductions compare rational estimates with thresholds. An additive-$\beta$ estimate $a$ satisfies $|a-M_T|\leq\beta$, $\beta\geq0$; an $\alpha$-approximation satisfies $J/\alpha\leq a\leq\alpha J$, $J>0$, $\alpha\geq1$. When certificates are included, they contain the Lyapunov polynomial, rational SOS lists and an explicit polynomial lower bound on $-\dot V$, positive at every nonzero state. Guarantees apply to every valid input. Approximation factors are bounded by a fixed polynomial in total input bit length $\mathcal L$, including proofs.

Existential real sentences assert feasibility of Boolean combinations of rational polynomial equalities and inequalities. $\exists\mathbb R$ consists of decision problems polynomial-time many-one reducible to this feasibility problem; $\forall\mathbb R$ consists of their complements. Completeness requires membership and hardness \cite{schaefer2017fixed}. Existential theory of the reals with inversion (ETR-INV) restricts variables to $[1/2,2]$ and equations to $x_i=1$, $x_i+x_j=x_k$ and $x_ix_j=1$; it is $\exists\mathbb R$-complete \cite{abrahamsen2018gallery} (\Cref{app:real-hardness}).

\section{Energy exchange and Lyapunov functions}\label{sec:model}\label{sec:equivalence}\label{sec:quartic}
For positive integers $d,m$, we use real symmetric $d\times d$ matrices $\mathcal Q=(Q_i)_{i=1}^m$, rational in computational inputs. The homogeneous quadratic map $q(x)=q_{\mathcal Q}(x)=(x^\top Q_i x)_{i=1}^m$ has common-root set $q^{-1}(0)$. Its roots become equilibria of
\begin{equation}\label{eq:field}
 \dot x=-Dq(x)^\top y,\qquad \dot y=2q(x)-y.
\end{equation}
Here $x\in\mathbb R^d$, $y\in\mathbb R^m$, and the full state $z=(x,y)\in\mathbb R^{d+m}$ has unrestricted initial coordinates. The field $F_q$ is quadratic since $Dq$ is linear. Energy cancellation ensures stability before we address convergence.

\begin{lem}\label[lemma]{lem:energy}
Every solution of system~\labelcref{eq:field} is forward complete with nonincreasing Euclidean norm. The origin is Lyapunov stable for every $q$. For the Euclidean energy $E(z)=\left\lVert z\right\rVert^2/2$, the energy derivative and linearization are
\begin{equation}\label{eq:energy}
 \dot E=-\left\lVert y\right\rVert^2,\qquad DF_q(0)=\operatorname{diag}(0_{d\times d},-I_m).
\end{equation}
Here $I_m$ is the $m\times m$ identity matrix.
\end{lem}
\begin{pf}
Local Lipschitz continuity gives unique local solutions. Euler's identity $Dq(x)x=2q(x)$ gives $x^\top\dot x=-2q(x)^\top y$. Adding $y^\top\dot y$ proves \Cref{eq:energy}. Invariant closed balls, where $F_q$ is bounded, give stability and forward completeness. The coupling's Jacobian vanishes at zero.
\qed\end{pf}

Energy alone does not imply attraction to the origin. We write $X=\lVert x\rVert^2$, $Y=\lVert y\rVert^2$, $C=\lVert q(x)\rVert^2$, $\rho=X+Y$ and $B=\sum_i\lVert Q_i\rVert_F^2$. On $y=0$, the energy derivative vanishes, whereas the correction $-y^\top q(x)$ has derivative $-2C$. The quartic coefficient $B+1$ ensures global positivity:
\begin{equation}\label{eq:quartic}
 V_4(x,y)=\frac12\rho+(B+1)\rho^2-y^\top q(x).
\end{equation}

\begin{thm}\label[theorem]{thm:quartic}
For every $q$, the degree-four polynomial in \Cref{eq:quartic} is positive definite and radially unbounded. Its positivity and dissipation bounds are
\begin{align}
 V_4&\geq\tfrac12X+\tfrac14Y+\rho^2,\label{eq:positive}\\
 \dot V_4&\leq-\tfrac12Y-\tfrac32C-4XY-4(B+1)Y^2.\label{eq:negative}
\end{align}
It is strict if and only if $q^{-1}(0)=\{0\}$. For rational $\mathcal Q$, both $V_4$ and $-\dot V_4$ admit explicit rational weighted SOS decompositions, as proved in \Cref{app:residuals}.
\end{thm}
\begin{pf}
Cauchy--Schwarz gives
\begin{equation}\label{eq:cauchy}
 C\leq BX^2,\qquad \left\lVert Dq(x)^\top y\right\rVert^2\leq4BXY.
\end{equation}
The bound on $C$ follows from $q_i(x)=\langle Q_i,xx^\top\rangle_F$ and $\lVert xx^\top\rVert_F=X$. We bound $\lVert Dq(x)^\top y\rVert^2$ by Cauchy--Schwarz, using $Dq(x)^\top y=2\sum_i y_iQ_i x$ and $\sum_i\lVert Q_i x\rVert^2\leq BX$. Using $|y^\top q(x)|\leq Y/4+C$ gives
\begin{equation}\label{eq:results-detail-1}
 V_4-\tfrac12X-\tfrac14Y\geq(B+1)\rho^2-BX^2\geq\rho^2.
\end{equation}
Positivity and radial unboundedness follow. Differentiating with $\dot\rho=-2Y$ gives
\begin{align}
 \frac{d}{dt}(y^\top q(x))={}&2C-y^\top q(x)-\lVert Dq(x)^\top y\rVert^2,\label{eq:results-detail-2}\\
 \dot V_4={}&-Y-4(B+1)\rho Y-2C+y^\top q(x)\notag\\
 &+\lVert Dq(x)^\top y\rVert^2.\label{eq:exact-derivative}
\end{align}
Combining \Cref{eq:cauchy} with $y^\top q(x)\leq(Y+C)/2$ gives \Cref{eq:negative}, strict away from zero when $q^{-1}(0)=\{0\}$. On the undamped slice,
\begin{equation}\label{eq:slice}
 \dot V_4(x,0)=-2\left\lVert q(x)\right\rVert^2.
\end{equation}
Nonzero roots prevent strictness.
\qed\end{pf}

The common-root condition also characterizes attraction: it identifies the trajectories on which energy dissipation vanishes.

\begin{thm}\label[theorem]{thm:equivalence}
For system~\labelcref{eq:field}, the following are equivalent:
\begin{equation}\label{eq:results-detail-3}
 q^{-1}(0)=\{0\}\ \Longleftrightarrow\ \mathrm{LAS}\ \Longleftrightarrow\ \mathrm{GAS}.
\end{equation}
The largest invariant subset of $\{\dot E=0\}$ is precisely $\{(x,0):q(x)=0\}$.
\end{thm}
\begin{pf}
Remaining in $y=0$ requires $q(x)=0$ by system~\labelcref{eq:field}; then $\dot x=0$ gives precisely these equilibria. For $q^{-1}(0)=\{0\}$, this subset is the origin. Compact, positively invariant energy sublevels (\Cref{lem:energy}) give global attraction by LaSalle's invariance principle \cite{lasalle1960extensions}; \Cref{lem:energy} gives stability, hence GAS. Conversely, homogeneity scales any nonzero root to equilibria arbitrarily near zero, preventing local attraction. GAS implies LAS.
\qed\end{pf}

We next determine the minimum degree of a strict global Lyapunov polynomial.

\begin{thm}[Degree]\label[theorem]{prop:degree}
When system~\labelcref{eq:field} is GAS, the minimum degree of a strict global polynomial Lyapunov function is exactly four. No strict quadratic Lyapunov function exists, even locally, for any system in this family.
\end{thm}
\begin{pf}
A nonzero leading cubic changes sign under $z\mapsto-z$, contradicting global positivity. A strict global Lyapunov polynomial of degree at most three therefore has the form $V(z)=z^\top Pz$, $P\succ0$, its constant and linear terms vanishing at its minimum. At $y=0$, its off-diagonal block $P_{xy}$ gives
\begin{equation}\label{eq:results-detail-4}
 \dot V(x,0)=4x^\top P_{xy}q(x).
\end{equation}
The derivative reverses sign under $x\mapsto-x$, precluding strict decrease at both $(x,0)$ and $(-x,0)$, even locally. The construction in \Cref{thm:quartic} attains degree four; $E$ is non-strict.
\qed\end{pf}

\section{Distance to loss of GAS and halving time}\label{sec:law}
The weakest unit-sphere coupling quantifies the coefficient distance to nonzero equilibria. We define
\begin{equation}\label{eq:sphere-residual}
 \gamma(q)=\min_{\lVert u\rVert=1}\lVert q(u)\rVert,\qquad
 C(x)\geq\gamma(q)^2 X^2.
\end{equation}
Compactness ensures that the minimum is attained; homogeneity gives the inequality. By \Cref{thm:equivalence}, $\gamma(q)>0$ characterizes GAS. For symmetric $P_i\in\mathbb R^{d\times d}$, we write $\mathcal P=(P_i)_{i=1}^m$ and $q_{\mathcal P}(x)=(x^\top P_ix)_i$, and measure coefficient changes by the Frobenius distance
\begin{equation}\label{eq:results-detail-20}
 \delta(\mathcal Q,\mathcal P)=\left(\sum_i\lVert P_i-Q_i\rVert_F^2\right)^{1/2}.
\end{equation}

\begin{thm}[Distance]\label[theorem]{prop:distance}
Over real symmetric tuples $\mathcal P$ for which system~\labelcref{eq:field} with $q=q_{\mathcal P}$ is not GAS, $\delta(\mathcal Q,\mathcal P)$ attains its minimum $\gamma(q)$. For every real symmetric tuple $\mathcal P$, we have $|\gamma(q_{\mathcal P})-\gamma(q)|\leq\delta(\mathcal Q,\mathcal P)$. For $R\geq0$, every system with $\delta(\mathcal Q,\mathcal P)\leq R$ is GAS if and only if $R<\gamma(q)$.
\end{thm}
\begin{pf}
Cauchy--Schwarz on $\langle P_i-Q_i,xx^\top\rangle_F$ gives
\begin{equation}\label{eq:perturbation-residual}
 \lVert q_{\mathcal P}(x)-q(x)\rVert\leq\delta(\mathcal Q,\mathcal P)\lVert x\rVert^2.
\end{equation}
Minimizing over unit $x$ and interchanging the tuples gives the absolute residual bound. A unit root of $q_{\mathcal P}$ requires $\delta(\mathcal Q,\mathcal P)\geq\gamma(q)$. At a unit minimizer $u$ in \Cref{eq:sphere-residual}, the rank-one perturbations
\begin{equation}\label{eq:rank-one-perturbation}
 P_i=Q_i-q_i(u)uu^\top
\end{equation}
give $q_{\mathcal P}(u)=0$ and, since $\lVert uu^\top\rVert_F=1$, $\delta(\mathcal Q,\mathcal P)^2=\sum_iq_i(u)^2=\gamma(q)^2$.
\qed\end{pf}

\begin{figure}[t]
\centering
\includegraphics[width=0.8\columnwidth]{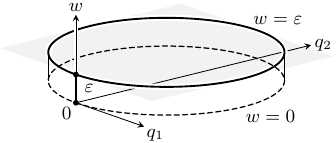}
\caption{Residual image and distance to loss of GAS. The unit-circle image of $q(x)=(x_1x_2,x_2^2)$ contains zero (dashed). Appending $\varepsilon\lVert x\rVert^2$ gives $\gamma(\widehat q)=\varepsilon$ (solid) and GAS; $\varepsilon=1/8$ (\Cref{prop:distance}).}
\label{fig:geometry}
\end{figure}

\noindent\textit{Remark.} For $m\geq d$ and a nonzero tuple, the condition-number theorem~\cite{burgisser2019homology} gives the same distance identity: the Weyl coefficient norm of $x^\top Qx$ equals $\lVert Q\rVert_F$. Our proof covers every positive $m$. The same perturbed $q$ enters both coupling equations; dimensions and damping $-y$ remain fixed. Thus the distance concerns systems with a nonzero equilibrium (\Cref{fig:geometry}).

At $t\geq0$, with solution map $\Phi_t$, the maximum retained fraction is
\begin{equation}\label{eq:results-detail-9}
 M_t=\max_{\left\lVert z_0\right\rVert=1}\left\lVert\Phi_t(z_0)\right\rVert^2.
\end{equation}
Compactness and continuous dependence give attainment. By \Cref{eq:energy}, the least dissipated fraction $1-M_t$ equals twice the unit-sphere minimum of $\int_0^t\lVert y\rVert^2\,ds$. Under GAS, $t>0$ gives $M_t<1$: equality forces $y=0$ on $[0,t]$ and hence a nonzero equilibrium. Since $M_t$ is nonincreasing, we define the worst-case halving time $T_{1/2}=\inf\{t\geq0:M_t\leq1/2\}$, with $\inf\emptyset=\infty$.

The residual bound strengthens quartic dissipation to yield decay from every unit initial state.

\begin{lem}[Uniform decay]\label[lemma]{thm:finite}
If $0<\mu\leq1/4$ and $\left\lVert q(x)\right\rVert^2\geq\mu\left\lVert x\right\rVert^4$ for every $x$, every solution with a unit initial condition satisfies $\rho(t)\leq32(B+1)^2/(\mu t)$ for all $t>0$.
\end{lem}
\begin{pf}
Unit initial states give $\rho(t)\leq1$. With $c_V=2(B+1)$, \Cref{eq:positive} gives $\rho/4\leq V_4$. Young's inequality gives $|y^\top q|\leq Y/4+C\leq Y/4+BX^2$, hence $V_4\leq3\rho/4+(2B+1)\rho^2\leq c_V\rho$. Since $Y\leq1$, $\rho^2\leq2X^2+2Y$. The residual bound gives $Y/2+3C/2\geq\mu(X^2+Y)\geq\mu\rho^2/2$. With $\lambda=\mu/(2c_V^2)$, \Cref{eq:negative} yields
\begin{align}
 \rho/4&\leq V_4\leq c_V\rho,\label{eq:results-detail-13}\\
 \dot V_4&\leq-\frac{\mu}{2}\rho^2\leq-\lambda V_4^2.\label{eq:results-detail-14}
\end{align}
For $\upsilon(t)=V_4(z(t))>0$, monotonicity permits integrating $(1/\upsilon)'\geq\lambda$ on $[0,t]$, giving $\upsilon(t)\leq1/(\lambda t)$. Hence $\rho(t)\leq4\upsilon(t)\leq4/(\lambda t)=32(B+1)^2/(\mu t)$. If $\upsilon(t)=0$, then $\rho(t)=0$.
\qed\end{pf}

Minimizing the coupling yields a slow trajectory; zero coupling gives an equilibrium.

\begin{lem}[Slow trajectory]\label[lemma]{lem:slowray}
Every system~\labelcref{eq:field} has a unit initial state whose solution satisfies $\lVert z(t)\rVert^2\geq1-8\gamma(q)^2t$ for all $t\geq0$.
\end{lem}
\begin{pf}
If $\gamma(q)=0$, a unit root $u$ gives the equilibrium $(u,0)$. Otherwise, we use a unit minimizer $u$ in \Cref{eq:sphere-residual} and set $\gamma=\gamma(q)$. Homogeneity makes $\lVert q(x)\rVert^2-\gamma^2\lVert x\rVert^4$ non-negative and zero at $u$, so its gradient vanishes at $u$:
\begin{equation}\label{eq:minimizer-stationarity}
 Dq(u)^\top q(u)=2\gamma^2 u.
\end{equation}
Since $q(ru)=r^2q(u)$ and $Dq(ru)=rDq(u)$, the substitutions $x=ru$ and $y=sq(u)$ solve system~\labelcref{eq:field} exactly when
\begin{equation}\label{eq:minimizing-ray}
 r'=-2\gamma^2 rs,\qquad s'=2r^2-s.
\end{equation}
By uniqueness, $x=ru$ and $y=sq(u)$ with $(r(0),s(0))=(1,0)$ yield the solution from $(u,0)$. The identity $(r^2+\gamma^2 s^2)'=-2\gamma^2 s^2$ bounds both coordinates, ensuring global continuation and $r^2\leq1$; integrating the second equation gives $0\leq s\leq2$. Hence $(r^2)'=-4\gamma^2 r^2s\geq-8\gamma^2$ and $\lVert z(t)\rVert^2\geq r(t)^2\geq1-8\gamma^2 t$.
\qed\end{pf}

These decay estimates give two-sided halving-time bounds in terms of the residual and hence the coefficient distance.

\begin{thm}[Halving time]\label[theorem]{thm:law}
For every real symmetric tuple $\mathcal Q$ and $t>0$, we set $\gamma=\gamma(q)$ and $\gamma_*=\min(\gamma,1/2)$. The maximum retained energy fraction satisfies
\begin{equation}\label{eq:law}
 1-8\gamma^2t\;\leq\;M_t\;\leq\;\frac{32(B+1)^2}{\gamma_*^2\,t},
\end{equation}
with right side $+\infty$ if $\gamma=0$. Hence $1/(16\gamma^2)\leq T_{1/2}\leq64(B+1)^2/\gamma_*^2$, interpreting both endpoints as $+\infty$ when $\gamma=0$. Thus $T_{1/2}<\infty$ if and only if system~\labelcref{eq:field} is GAS.
\end{thm}
\begin{pf}
\Cref{lem:slowray} gives the lower bound; for $\gamma>0$, \Cref{thm:finite} with $\mu=\gamma_*^2$ gives the upper bound. The halving-time endpoints follow by comparison with $1/2$. If $\gamma=0$, a unit equilibrium gives $M_t=1$ for all $t$.
\qed\end{pf}

\noindent\textit{Remark.} For $0<\gamma\leq1/2$, the upper halving-time bound is $1024(B+1)^2$ times the lower bound. By \Cref{prop:distance}, $\gamma$ is the distance to loss of GAS.

\section{Complexity of exact stability}\label{sec:complexity}\label{sec:real}

Before studying performance under known stability, we establish the difficulty of deciding stability itself. We adapt Ahmadi's quartic encoding \cite{ahmadi2012difficulty} from ONE-IN-THREE satisfiability, which is NP-complete for unnegated literals with repeats \cite{schaefer1978complexity}. Each clause requires exactly one true literal occurrence, counted with multiplicity.

The Boolean input lists a binary variable count and literal triples with binary identifiers. For $\varphi$ with $n\geq1$ variables $b_i$ and $c\geq0$ clauses, we allow negations and repeats. Removing unused variables and relabeling, retaining one if none occurs, gives $n\leq\max(1,3c)$. Homogenizing with $h$, for $x=(h,a)$, $a\in\mathbb R^n$, we encode $a_i=0$ or $h$ by
\begin{equation}\label{eq:bool}
 q_i(x)=a_i(a_i-h),\qquad i=1,\ldots,n.
\end{equation}
Literal $k$ of clause $j$ is $\ell_{jk}$, with encoded values $L(b_i)=a_i$ and $L(\neg b_i)=h-a_i$. Clause $j$ requires the encoded values to sum to $h$. Multiplication by $h$ preserves quadratic homogeneity, giving
\begin{equation}\label{eq:clause}
 q_{n+j}(x)=h\left(\sum_{k=1}^3 L(\ell_{jk})-h\right).
\end{equation}
The map $q_\varphi$ turns satisfying assignments into nonzero roots while excluding spurious roots with $h=0$.

\begin{lem}\label[lemma]{lem:encoding}
The formula $\varphi$ has exactly one true literal per clause under some assignment if and only if the map in \Cref{eq:bool,eq:clause} has a nonzero real common root.
\end{lem}
\begin{pf}
A satisfying assignment $b\in\{0,1\}^n$ gives the nonzero root $(1,b)$. Conversely, $h=0$ forces every $a_i=0$, excluding a nonzero root. Dividing the root equations in \Cref{eq:bool,eq:clause} by $h^2$ gives $a_i/h\in\{0,1\}$ and encoded literal values summing to one.
\qed\end{pf}

The reduction maps unsatisfiability to asymptotic stability; uniform coefficient bounds make the hardness strong.

\begin{thm}\label[theorem]{thm:hardness}
Deciding whether system~\labelcref{eq:field} is GAS, or LAS, is strongly coNP-hard under polynomial-time many-one reductions, even with integer field coefficients of magnitude at most six. Although $V_4$ is constructed in polynomial time, deciding its strictness is coNP-hard.
\end{thm}
\begin{pf}
With $d=n+1$, $m=n+c$ and total dimension $N=d+m=2n+c+1$, \Cref{lem:encoding,thm:equivalence} give the equivalence:
\begin{equation}\label{eq:results-detail-5}
 \begin{aligned}
 \varphi\text{ unsatisfiable}
 &\Longleftrightarrow 0\text{ is LAS for }F_{q_\varphi}\\
 &\Longleftrightarrow 0\text{ is GAS for }F_{q_\varphi}.
\end{aligned}
\end{equation}
Complementing the NP-complete source gives strong coNP-hardness: uniform integer bounds persist under unary coefficient encoding. Each collected clause polynomial is $\nu_0h^2+\sum_i \nu_iha_i$, where $\nu_0\in\{-1,0,1,2\}$ and $\sum_i|\nu_i|\leq3$. Variable coefficients are $1,-1$. Doubling and differentiation give integer field coefficients bounded by six; distinct $y_i$ prevent accumulation across constraints. At most $6n+12c$ monomial terms occur. Splitting cross-term coefficients equally between symmetric entries gives half-integer matrices $Q_i$. Variable-constraint matrices have squared Frobenius norm $3/2$; clause matrices have $\nu_0^2+\tfrac12\sum_i \nu_i^2\leq17/2$, giving $B\leq(3n+17c)/2$. Computing $B$ uses $O(N)$ squares of bounded rational entries. After collecting monomials, $V_4$ has at most $N^2+5N$ terms, with half-integer coefficients of magnitude at most $3n+17c+5$. Its coefficients and indices use $O(\log N)$ bits. Fixed-depth coordinate loops construct the field and quartic in polynomial bit time \cite{hoeven2013bit}. \Cref{thm:quartic} identifies strictness with absence of nonzero common roots.
\qed\end{pf}

We next prove that stability verification is complete for $\forall\mathbb R$. Quadratic feasibility is NP-hard \cite{hillar2013most} and $\exists\mathbb R$-complete \cite{schaefer2017fixed}, also for nonzero homogeneous roots \cite{schaefer2018tensor}. By \Cref{thm:equivalence}, stability has the complementary classification.

\begin{thm}\label[theorem]{thm:real}
The following properties are $\forall\mathbb R$-complete for rational symmetric matrix inputs: local asymptotic stability of system~\labelcref{eq:field}, global asymptotic stability, and strictness of the quartic function in \Cref{eq:quartic}. Hardness holds with integer field coefficients of magnitude at most six. The complement properties are $\exists\mathbb R$-complete.
\end{thm}
\begin{pf}
By \Cref{thm:quartic,thm:equivalence}, each complement asks for a nonzero common root. Homogeneity permits the unit constraints $q(x)=0$, $\left\lVert x\right\rVert^2=1$, of polynomial rational description length. Root existence thus belongs to $\exists\mathbb R$, stability and strictness to $\forall\mathbb R$. For hardness, \Cref{app:real-hardness} reduces ETR-INV feasibility to nonzero common roots with integer field coefficients of magnitude at most six, making every complement property $\exists\mathbb R$-hard.
\qed\end{pf}

The residual-distance identity also classifies closed coefficient balls, even with GAS centers and a fixed query radius.

\begin{cor}\label[corollary]{cor:ball}
Given a rational symmetric tuple $\mathcal Q$ and a rational radius $R\geq0$, it is $\forall\mathbb R$-complete to decide whether system~\labelcref{eq:field} is GAS for every real symmetric $\mathcal P$ with $\delta(\mathcal Q,\mathcal P)\leq R$, even at radius one around centers with proved GAS and integer field coefficients of magnitude at most six.
\end{cor}
\begin{pf}
We append $\lVert x\rVert^2$ to the map in \Cref{thm:real}, making the center GAS; the queried radius is one. For $q^+=(q,\lVert x\rVert^2)$,
\begin{equation}\label{eq:unit-radius-reduction}
 \gamma(q^+)^2=1+\gamma(q)^2.
\end{equation}
By \Cref{prop:distance}, its closed unit ball consists of GAS systems if and only if $q$ has no nonzero common root, so \Cref{thm:real} gives hardness. Every center has $\gamma(q^+)\geq1$; added field coefficients are at most two. Failure has an existential witness, a symmetric tuple $\mathcal P$ and unit vector $u$ satisfying $\sum_i\lVert P_i-Q_i\rVert_F^2\leq R^2$ and $u^\top P_i u=0$ for every $i$. These polynomial constraints give $\exists\mathbb R$ membership for the complement.
\qed\end{pf}

\section{Inapproximability of dissipation estimates}\label{sec:examples}
We first quantify the residual gap between satisfiable and unsatisfiable formulas.

\begin{figure}[t]
\centering
\includegraphics[width=\columnwidth]{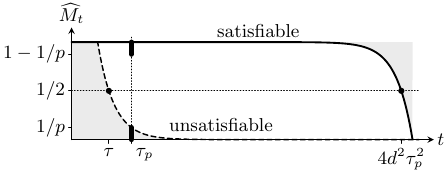}
\caption{Schematic bounds from \Cref{thm:allstable}. Satisfiable inputs retain at least $1-8\varepsilon^2t$ (solid); unsatisfiable inputs retain at most $\tau/(2t)$ (dashed). Here $\tau=2^{18}$, $p=8$, $d=2$, $\varepsilon=2^{-24}$; time is logarithmic. Thick bars show retained-fraction bounds at $\tau_p$; dots mark halving-time bounds.}
\label{fig:performance}
\end{figure}

\begin{lem}[Residual lower bound]
For an unsatisfiable $n$-variable formula, its encoded map satisfies\label[lemma]{lem:gap}
\begin{equation}\label{eq:results-detail-10}
 \left\lVert q(x)\right\rVert^2\geq\mu_0\left\lVert x\right\rVert^4,\qquad
 \mu_0=[8(n+1)]^{-2}.
\end{equation}
\end{lem}
\begin{pf}
Assuming otherwise and normalizing $\lVert x\rVert=1$, we use $x=(h,a)$, $A=\lVert a\rVert^2$, $\kappa=\lVert q(x)\rVert<1/[8(n+1)]$. Cauchy--Schwarz gives \Cref{eq:results-detail-11}; then $A\leq|h|\sqrt{nA}+\sqrt n\,\kappa\leq(A+nh^2)/2+\sqrt n\,\kappa$ and $A+h^2=1$ yield \Cref{eq:results-detail-12}.
\begin{align}
 A-h\sum_i a_i&\leq\sqrt n\,\kappa,\label{eq:results-detail-11}\\
 1&\leq(n+1)h^2+2\sqrt n\,\kappa.\label{eq:results-detail-12}
\end{align}
Since $2\sqrt n\leq n+1$, the threshold gives $h^2>7/[8(n+1)]>0$. Dividing by $h$, we set $\zeta_i=a_i/h$ and $\theta=\kappa/h^2<1/7$. Each variable satisfies $|\zeta_i(\zeta_i-1)|\leq\theta$.

Since $\max(|\zeta_i|,|\zeta_i-1|)\geq1/2$, rounding to the nearest $b_i\in\{0,1\}$ gives $|\zeta_i-b_i|\leq2|\zeta_i(\zeta_i-1)|\leq2\theta$. Normalized clause errors are at most $\theta$; rounding three literals changes each sum by at most $6\theta$, including negations and repeats. Rounded integer clause errors have magnitude at most $7\theta<1$, enforcing every clause and contradicting unsatisfiability. Thus $\kappa^2\geq\mu_0$.
\qed\end{pf}

\noindent\textit{Remark.} Sharpness up to a constant factor follows from the unsatisfiable clause $(b_1,b_1,b_1)$ and clauses $(\neg b_i,\neg b_i,b_i)$ forcing $b_i=1$ for $i>1$. The point $h=1$, $a_1=0$, $a_i=1$ for $i>1$ gives $\lVert q(x)\rVert^2/\lVert x\rVert^4=1/n^2$.

Appending $\varepsilon\lVert x\rVert^2$ removes nonzero equilibria while preserving the residual gap, separating satisfiable and unsatisfiable inputs at a common time (\Cref{fig:performance}).

\begin{thm}[Tight inapproximability]\label[theorem]{thm:allstable}
For a formula $\varphi$ in ONE-IN-THREE, we write $q=q_\varphi$, $B=\sum_i\lVert Q_i\rVert_F^2$ and $d=n+1$. Integer $p\geq2$ is fixed for additive hardness or equals $\max(2,\ell^{r_0})$, with fixed integer $r_0\geq0$ and binary formula length $\ell$. With $\tau=64(B+2)^2/\mu_0$ and $K=\lceil p/2\rceil$, we evaluate at $\tau_p=K\tau$ after appending the radial term with coefficient $\varepsilon=1/(8d\tau_p)$ to form $\widehat q=(q,\varepsilon\lVert x\rVert^2)$. Its maximum retained fraction is $\widehat M_t$. All systems are GAS, with rational field coefficients of magnitude at most six and denominators dividing $4d\tau_p$. Strict quartic Lyapunov functions and rational SOS proofs have polynomial-time constructions. If $\varphi$ is satisfiable, then
\begin{equation}\label{eq:tight-sat}
 \gamma(\widehat q)=\varepsilon,\quad \widehat M_{\tau_p}\geq1-\tfrac1p,\quad T_{1/2}\geq4d^2\tau_p^2,
\end{equation}
and if $\varphi$ is unsatisfiable, then
\begin{equation}\label{eq:tight-unsat}
 \gamma(\widehat q)\geq\sqrt{\mu_0},\quad \widehat M_{\tau_p}\leq\tfrac1p,\quad T_{1/2}\leq\tau.
\end{equation}
Hence additive-$\beta$ estimation of $\widehat M_{\tau_p}$ is NP-hard for every fixed rational $0<\beta<1/2-1/p$. Choosing $p$ covers every fixed rational $0<\beta<1/2$; the constant estimate $1/2$ attains error $\beta=1/2$. For factor $\alpha=\alpha(\mathcal L)$, $\alpha$-approximation is NP-hard for $T_{1/2}$ when $\alpha^2<4d^2K^2\tau$, for $\gamma(\widehat q)$ when $\alpha^2<\tau_p$, and for $1-\widehat M_{\tau_p}$ when $\alpha^2<4d^2\tau_p$. Each bound must hold on all constructed inputs.
\end{thm}
\begin{pf}
The radial component gives the squared residual:
\begin{equation}\label{eq:results-detail-16}
 \left\lVert\widehat q(x)\right\rVert^2=\left\lVert q(x)\right\rVert^2+\varepsilon^2\left\lVert x\right\rVert^4.
\end{equation}
Thus $\gamma(\widehat q)\geq\varepsilon>0$. On every input, the quartic and SOS constructions in \Cref{thm:quartic,app:residuals} give $-\dot V_4\geq Y/2+3\varepsilon^2X^2/2$. The lists and lower bound certify strictness and GAS without deciding $\varphi$. The augmented tuple satisfies $\widehat B=B+d\varepsilon^2\leq B+1$.

For satisfiable $\varphi$, a unit root gives $\widehat q(u)=(0,\varepsilon)$ and $\gamma(\widehat q)=\varepsilon$. \Cref{lem:slowray} gives $\widehat M_{\tau_p}\geq1-1/(8d^2\tau_p)\geq1-1/p$, since $\tau_p\geq p/2$. \Cref{thm:law} gives $T_{1/2}\geq1/(16\varepsilon^2)=4d^2\tau_p^2$.

For unsatisfiable $\varphi$, \Cref{lem:gap} gives $\gamma(\widehat q)^2\geq\mu_0$. Hence \Cref{thm:finite,thm:law} give $\widehat M_{\tau_p}\leq32(B+2)^2/(\mu_0\tau_p)=1/(2K)\leq1/p$ and $T_{1/2}\leq\tau$.

For $0<\beta<1/2-1/p$, comparing estimates with $1/2$ decides satisfiability. The halving-time bounds have separation ratio $4d^2K^2\tau$. For the residual bounds this ratio is $\tau_p$, since $\sqrt{\mu_0}=1/(8d)$. Satisfiable inputs dissipate at most $1/(8d^2\tau_p)$ and unsatisfiable inputs at least $1/2$, giving ratio $4d^2\tau_p$. For rational $0<L<H$ and $\alpha^2<H/L$, comparing the squared estimate with $HL$ separates the cases without computing $\alpha$.

Since $2B\in\mathbb Z$, $\tau$ and $\tau_p$ are integers, polynomially bounded in $\ell$ for each fixed $p$ or $r_0$. Added coefficients have magnitude at most $2\varepsilon$ and denominators dividing $4d\tau_p$; all others are integers. Relabeling gives $n+c=O(\ell)$. The field, quartic and SOS lists (\Cref{app:residuals}) take polynomial time and have length $\mathcal L\leq C_0(\ell+\log p)^{s_0}$ for fixed constants; only numeric bit lengths depend on $p$. For $\alpha(\mathcal L)\leq C_1(1+\mathcal L)^{a_0}$, choosing $p=\max(2,\ell^{r_0})$ with fixed integer $r_0>2a_0s_0$ gives $\alpha(\mathcal L)^2=o(p)$. Every separation ratio is at least $p/2$; finitely many smaller formulas are decided directly.
\qed\end{pf}

The distance identity guarantees GAS in the coefficient ball. The next result concerns rational coefficient-tuple inputs, with stability promised by this ball.

\begin{thm}[Coefficient perturbations]\label[theorem]{cor:robust}
The real symmetric tuple $\widetilde{\mathcal Q}=(\widetilde Q_i)_{i=1}^{m+1}$ perturbs the tuple in \Cref{thm:allstable}, with $\delta(\widehat{\mathcal Q},\widetilde{\mathcal Q})\leq\varepsilon/2$. For $\widetilde q=q_{\widetilde{\mathcal Q}}$ and $\widetilde B=\sum_i\lVert\widetilde Q_i\rVert_F^2$, its maximum retained fraction is $\widetilde M_t$. Every such system is GAS. On satisfiable inputs $\widetilde M_{\tau_p}\geq1-9/(32d^2\tau_p)$; on unsatisfiable inputs $\widetilde M_{\tau_p}\leq4/(7K)$. For each fixed polynomial-time rule mapping source formulas to rational tuples $\widetilde{\mathcal Q}$ in this ball, additive-$\beta$ estimation at $\tau_p$ is NP-hard for every fixed rational $0<\beta<1/2-1/p$.
\end{thm}
\begin{pf}
By \Cref{prop:distance}, $|\gamma(\widetilde q)-\gamma(\widehat q)|\leq\varepsilon/2$, so $\gamma(\widetilde q)\geq\varepsilon/2>0$ giving GAS. On satisfiable inputs, $\gamma(\widetilde q)\leq3\varepsilon/2$, and \Cref{lem:slowray} gives $\widetilde M_{\tau_p}\geq1-18\varepsilon^2\tau_p=1-9/(32d^2\tau_p)$.

On unsatisfiable inputs, $\varepsilon/2\leq\sqrt{\mu_0}/16$ because $\tau_p\geq8$, so $\gamma(\widetilde q)^2\geq(15/16)^2\mu_0\geq7\mu_0/8$. Since $\sqrt{\widehat B}\leq\sqrt B+\sqrt d\,\varepsilon$, the triangle inequality and $\sqrt d+1/2\leq2d$ give $\sqrt{\widetilde B}\leq\sqrt B+2d\varepsilon$. Squaring gives $\widetilde B-B\leq\sqrt B/(2\tau_p)+1/(16\tau_p^2)\leq1$, hence $\widetilde B\leq B+1$. \Cref{thm:finite} with $\mu=7\mu_0/8$ gives $\widetilde M_{\tau_p}\leq32(B+2)^2/((7\mu_0/8)\tau_p)=4/(7K)$.

The fixed rational perturbation rule composes in polynomial time with the Boolean construction. The satisfiable bound on $1-\widetilde M_{\tau_p}$ and the unsatisfiable bound on $\widetilde M_{\tau_p}$ sum to at most $(8/7+9/128)/p<2/p$, because $\tau_p\geq8K$ and $d\geq1$. The retained-fraction gap exceeds $1-2/p>2\beta$; thresholding an additive-$\beta$ estimate at $4/(7K)+\beta$ decides satisfiability.
\qed\end{pf}

Direct damping makes the Euclidean energy a strict Lyapunov function and the Jacobian Hurwitz while retaining the performance gap.

\begin{thm}[Damping]\label[theorem]{prop:damping}
Using the Boolean map $q$ and $p,K$ from \Cref{thm:allstable}, we set $T_D=64(B+1)^2/\mu_0$, $T_p=KT_D$ and $\eta=1/(2pT_p)$, and replace the $x$-equation of system~\labelcref{eq:field} by $\dot x=-Dq^\top y-\eta x$. All systems are globally exponentially stable, with strict quadratic Lyapunov function $E$ and Hurwitz Jacobian. The maximum retained fraction at $T_p$ equals $e^{-1/p}\geq1-1/p$ on satisfiable inputs and is at most $1/(2K)\leq1/p$ on unsatisfiable inputs. Hence additive-$\beta$ estimation is NP-hard for every fixed rational $0<\beta<1/2-1/p$, with rational coefficients of magnitude at most six whose denominators divide $2pT_p$.
\end{thm}
\begin{pf}
Since $0<\eta\leq1$, the energy identity gives
\begin{equation}\label{eq:damped-energy}
 \dot E=-\eta X-Y\leq-2\eta E,\qquad \rho(t)\leq e^{-2\eta t}\rho(0).
\end{equation}
This proves completeness and exponential stability; the Jacobian $\operatorname{diag}(-\eta I_d,-I_m)$ is Hurwitz. Using $2y^\top q\leq X/4+4BXY$, the added term in $\dot V_4$ satisfies
\begin{equation}\label{eq:damped-quartic}
 \begin{aligned}
 &\eta[-X-4(B+1)\rho X+2y^\top q(x)]\\
 &\quad\leq-\eta[3X/4+4(B+1)X^2+4XY]\leq0.
 \end{aligned}
\end{equation}
The added term in $\dot V_4$ is nonpositive. The invariant unit ball therefore admits the decay estimate proved in \Cref{thm:finite}. Together with \Cref{lem:gap}, this gives $\rho(T_p)\leq32(B+1)^2/(\mu_0KT_D)=1/(2K)$ for all unit initial states on unsatisfiable inputs. Satisfiable inputs have a unit root $u$; $x(t)=e^{-\eta t}u$ and $y(t)=0$ attain \Cref{eq:damped-energy}, giving maximum $e^{-2\eta T_p}=e^{-1/p}$. Only the added linear coefficients have denominators; $T_p$ is a polynomially bounded integer.
\qed\end{pf}

Under $s=\eta t$, $dE/ds\leq-2E$ and $s(T_p)=1/(2p)$, constant for fixed $p$. The integer factor $\eta^{-1}=2pT_p$ scales the Boolean coefficients and changes added damping to $-1$, leaving polynomially bounded integer coefficients.

Higher auxiliary degree can sharpen convex finite-time bounds \cite{fantuzzi2020bounding}. Every fixed degree bound at least two includes our quadratic examples, with unchanged energy, initial sphere and certificates.

\begin{thm}[Polynomial systems]\label[theorem]{lem:extension}
For fixed $k\geq2$, class $\mathcal F_k$ comprises rational polynomial fields $F$ on $\mathbb R^N$, with $N\geq1$, degree at most $k$ and GAS origin. We use $M_t(F)$ from \Cref{eq:results-detail-9} for $\dot z=F(z)$, using a supremum, and $T_{1/2}(F)=\inf\{t\geq0:M_t(F)\leq1/2\}$ as first halving time because $M_t(F)$ need not be monotone. On $\mathcal F_k$, additive-$\beta$ estimation of $M_T(F)$ is NP-hard for every fixed rational $0<\beta<1/2$. Approximating $T_{1/2}(F)$, or $1-M_T(F)$ on inputs with $M_T(F)<1$, within any polynomial factor is NP-hard. These claims hold even with strict quartic Lyapunov proofs in the input. Additive hardness also covers globally exponentially stable fields with a strict quadratic Lyapunov function.
\end{thm}
\begin{pf}
Fields in \Cref{thm:allstable,prop:damping} already have degree at most $k$, with unchanged coordinates, time, energy, initial sphere and proofs. Here $M_T<1$ makes dissipated fractions positive, and \Cref{thm:law} gives $0<T_{1/2}<\infty$. The same reductions apply.
\qed\end{pf}

\section{Conclusion}\label{sec:discussion}
We study system~\labelcref{eq:field} and its directly damped variants in growing dimension. The objective uses full-state Euclidean energy and allows independent auxiliary initial conditions. Quadratic fluid and plasma models motivate this energy balance \cite{kaptanoglu2021promoting}. Auxiliary-function bounds \cite{fantuzzi2020bounding} seek to certify energy halving for every unit initial state by proving $M_T\leq1/2$ at time $T$.

Even with strict Lyapunov proofs in the input, estimating $M_T$ within any fixed rational additive error $0<\beta<1/2$ is NP-hard. Thus hierarchy convergence cannot guarantee polynomial worst-case runtime at this accuracy unless $\mathrm{P}=\mathrm{NP}$ (\Cref{lem:extension}).

\noindent\textbf{Future work.} We seek sharper dependence on $B$ in the halving-time bounds and conditions on $q$ permitting polynomial-time $M_T$ approximation with explicit error bounds.

\appendices
\numberwithin{equation}{section}
\section{Residual sum-of-squares identities}\label[appendix]{app:residuals}
The square expansions for \Cref{thm:quartic} are
\begin{align}
 &V_4=\tfrac12X+\tfrac14Y+\rho^2+\lVert y/2-q(x)\rVert^2\notag\\
 &\quad+B(\rho^2-X^2)+(BX^2-C),\label{eq:sos-positive}\\
 &-\dot V_4=\tfrac12Y+\tfrac32C+4(B+1)Y^2+\tfrac12\lVert y-q(x)\rVert^2\notag\\
 &\quad+4XY+4BXY-\lVert Dq(x)^\top y\rVert^2.\label{eq:sos-negative}
\end{align}
Since $\rho^2-X^2=2XY+Y^2$ and $XY=\sum_{r,l}(x_ry_l)^2$, all but the residuals $BX^2-C$ and $4BXY-\lVert Dq(x)^\top y\rVert^2$ are weighted SOS. Indices satisfy $1\leq i,l\leq m$ and $1\leq r\leq d$ unless shown otherwise. Vectorizing $Q_i$ and $xx^\top$ identically gives $v_i\in\mathbb R^{d^2}$ and $w\in\mathbb R^{d^2}$. Then $v_i^\top w=q_i(x)$, $\lVert v_i\rVert^2=\lVert Q_i\rVert_F^2$ and $\lVert w\rVert^2=X^2$, so Lagrange's identity gives
\begin{equation}\label{eq:sos-lagrange}
 BX^2-C=\sum_i\sum_{1\leq j<k\leq d^2}(v_{ij}w_k-v_{ik}w_j)^2.
\end{equation}
For the Jacobian residual, we define $\psi_{ir}=(Q_ix)_r$ and
\begin{equation}\label{eq:sos-minor}
 \Lambda_{irjk}=(Q_i)_{rj}x_k-(Q_i)_{rk}x_j.
\end{equation}
As $Dq(x)^\top y=2\sum_i y_iQ_ix$, adding and subtracting $4Y\allowbreak\sum_i\lVert Q_ix\rVert^2$ gives
\begin{align}\label{eq:sos-split}
 &4BXY-\lVert Dq(x)^\top y\rVert^2\notag\\
 &=4Y\sum_i\bigl(\lVert Q_i\rVert_F^2X-\lVert Q_ix\rVert^2\bigr)\notag\\
 &\quad+4\left(Y\sum_i\lVert Q_ix\rVert^2-\left\lVert\sum_i y_iQ_ix\right\rVert^2\right).
\end{align}
Applying Lagrange's identity rowwise to $Q_i$ and $x$, then to $y$ and $(\psi_{ir})_{i=1}^m$ for each $r$, expands \Cref{eq:sos-split}. With $Y=\sum_l y_l^2$, we obtain
\begin{align}
 &4BXY-\left\lVert Dq(x)^\top y\right\rVert^2
 =4\sum_l\sum_{i,r}\sum_{1\leq j<k\leq d}(y_l\Lambda_{irjk})^2\notag\\
 &\quad+4\sum_r\sum_{1\leq i<j\leq m}(y_i\psi_{jr}-y_j\psi_{ir})^2.\label{eq:sos-jacobian}
\end{align}
The Lagrange residual has $m\binom{d^2}{2}$ squares; the Jacobian residual has $m^2d\binom d2+d\binom m2$. For these pair-indexed residuals, literal lists retain all ordered pairs with half the displayed weights. Diagonal squares vanish, and pairing reversed indices recovers the unordered sums. Fixed-depth loops construct the lists in polynomial bit time using exact rational arithmetic. For fixed $p$ and fixed-$r_0$ in \Cref{thm:allstable}, intermediate integers are polynomially bounded in $\ell$; Euclidean fraction reduction preserves this time bound \cite{hoeven2013bit}. Empty sums vanish, and the identities establish non-negativity. For $\widehat q$ in \Cref{thm:allstable}, $V_4$ uses $\widehat q$ in its cross-term, $\widehat B$ in its quartic coefficient and the enlarged $y$ block. Here $\widehat C=\lVert\widehat q(x)\rVert^2=C+\varepsilon^2X^2$. Substituting these quantities in \Cref{eq:sos-negative} gives
\begin{equation}\label{eq:sos-strict-augmented}
 -\dot V_4\geq\tfrac12Y+\tfrac32\varepsilon^2X^2>0\quad(z\ne0).
\end{equation}

\section{Reduction from bounded real equations}\label[appendix]{app:real-hardness}
For \Cref{thm:real}, we use ETR-INV with $n\geq1$ variables $\chi_i\in[1/2,2]$ and $n_{\mathrm{eq}}$ equations of the forms
\begin{equation}\label{eq:results-detail-6}
 \chi_i=1,\qquad \chi_i+\chi_j=\chi_\ell,\qquad \chi_i\chi_j=1.
\end{equation}
The source lists its variables explicitly \cite{abrahamsen2018gallery}. Thus unary $n$ has polynomial description length; equation indices remain binary, and we retain all declared variables. With $x=(h,\xi,\sigma,\omega)$ and $\xi,\sigma,\omega\in\mathbb R^n$, we encode $(2\chi_i-1)(2-\chi_i)\geq0$ using a squared auxiliary variable. Keeping doubled coefficients within six, we impose $\omega_i=\xi_i$ and replace $5h\xi_i$ by $3h\xi_i+2h\omega_i$:
\begin{align}
 g_i&=-2\xi_i^2+3h\xi_i+2h\omega_i-2h^2-2\sigma_i^2,\label{eq:box-interval}\\
 f_i&=(\xi_i-\omega_i)^2.\label{eq:box-equality}
\end{align}
The components of $q$ are $g_i$, $f_i$ and the homogenized source polynomials
\begin{equation}\label{eq:real-source}
 h(\xi_i-h),\quad h(\xi_i+\xi_j-\xi_\ell),\quad \xi_i\xi_j-h^2.
\end{equation}
At a common root, $f_i=0$ forces $\omega_i=\xi_i$. If $h=0$, then $g_i=-2\xi_i^2-2\sigma_i^2=0$ forces $x=0$. At nonzero roots, division by $h^2$ gives
\begin{equation}\label{eq:results-detail-7}
 (2\chi_i-1)(2-\chi_i)=2(\sigma_i/h)^2,\qquad \chi_i=\xi_i/h.
\end{equation}
Thus $\chi_i\in[1/2,2]$. Dividing \Cref{eq:real-source} by $h^2$ recovers the source equations.
Conversely, source solutions give nonzero real roots with $h=1$, $\xi_i=\omega_i=\chi_i$ and $\sigma_i=((2\chi_i-1)(2-\chi_i)/2)^{1/2}$. Only this real witness uses square roots; constructed coefficients remain rational. Thus infeasibility maps exactly to GAS and quartic strictness, with $d=3n+1$ and $m=2n+n_{\mathrm{eq}}$.

The quadratic components have monomial coefficients of magnitude at most three and square-term coefficients of magnitude at most two, even for repeated indices. Doubling and differentiation give field coefficients of magnitude at most six; distinct $y_i$ separate contributions from different constraints in $\dot x$.
The $g_i$ matrix has squared Frobenius norm $37/2$; the $f_i$ matrix has squared norm $4$. Each source-equation matrix has squared norm at most three, giving $B\leq45n/2+3n_{\mathrm{eq}}$. Each component has at most five monomials. Fixed-depth loops over equation and matrix indices construct the tuple in polynomial bit time.


\begin{thebibliography}{10}
\providecommand{\url}[1]{#1}
\csname url@samestyle\endcsname
\providecommand{\newblock}{\relax}
\providecommand{\bibinfo}[2]{#2}
\providecommand{\BIBentrySTDinterwordspacing}{\spaceskip=0pt\relax}
\providecommand{\BIBentryALTinterwordstretchfactor}{4}
\providecommand{\BIBentryALTinterwordspacing}{\spaceskip=\fontdimen2\font plus
\BIBentryALTinterwordstretchfactor\fontdimen3\font minus
  \fontdimen4\font\relax}
\providecommand{\BIBforeignlanguage}[2]{{
\expandafter\ifx\csname l@#1\endcsname\relax
\typeout{** WARNING: IEEEtran.bst: No hyphenation pattern has been}
\typeout{** loaded for the language `#1'. Using the pattern for}
\typeout{** the default language instead.}
\else
\language=\csname l@#1\endcsname
\fi
#2}}
\providecommand{\BIBdecl}{\relax}
\BIBdecl

\bibitem{kaptanoglu2021promoting}
A.~A. Kaptanoglu, J.~L. Callaham, A.~Aravkin, C.~J. Hansen, and S.~L. Brunton,
  ``Promoting global stability in data-driven models of quadratic nonlinear
  dynamics,'' \emph{Physical Review Fluids}, vol.~6, no.~9, 2021,
  \href{https://doi.org/10.1103/PhysRevFluids.6.094401}{\nolinkurl{doi:10.1103/PhysRevFluids.6.094401}}.

\bibitem{fantuzzi2020bounding}
G.~Fantuzzi and D.~Goluskin, ``Bounding extreme events in nonlinear dynamics
  using convex optimization,'' \emph{SIAM Journal on Applied Dynamical
  Systems}, vol.~19, no.~3, 2020,
  \href{https://doi.org/10.1137/19M1277953}{\nolinkurl{doi:10.1137/19M1277953}}.

\bibitem{miller2026crash}
J.~Miller and M.~Sznaier, ``Quantifying the safety of trajectories by the data
  corruption needed to crash,'' \emph{SIAM Journal on Control and
  Optimization}, vol.~64, no.~4, 2026,
  \href{https://doi.org/10.1137/24M166262X}{\nolinkurl{doi:10.1137/24M166262X}}.

\bibitem{malisoff2009constructions}
M.~Malisoff and F.~Mazenc, \emph{Constructions of Strict {Lyapunov}
  Functions}.\hskip 1em plus 0.5em minus 0.4em\relax Springer, 2009,
  \href{https://doi.org/10.1007/978-1-84882-535-2}{\nolinkurl{doi:10.1007/978-1-84882-535-2}}.

\bibitem{mazenc2004strong}
F.~Mazenc and D.~Ne{\v{s}}i{\'c}, ``Strong {Lyapunov} functions for systems
  satisfying the conditions of {La Salle},'' \emph{IEEE Transactions on
  Automatic Control}, vol.~49, no.~6, 2004,
  \href{https://doi.org/10.1109/TAC.2004.829640}{\nolinkurl{doi:10.1109/TAC.2004.829640}}.

\bibitem{goyal2025guaranteed}
P.~Goyal, I.~Pontes~Duff, and P.~Benner, ``Guaranteed stable quadratic models
  and their applications in {SINDy} and operator inference,'' \emph{Physica D:
  Nonlinear Phenomena}, vol. 483, 2025,
  \href{https://doi.org/10.1016/j.physd.2025.134893}{\nolinkurl{doi:10.1016/j.physd.2025.134893}}.

\bibitem{ahmadi2012difficulty}
A.~A. Ahmadi, ``On the difficulty of deciding asymptotic stability of cubic
  homogeneous vector fields,'' in \emph{2012 American Control
  Conference}.\hskip 1em plus 0.5em minus 0.4em\relax IEEE, 2012,
  \href{https://doi.org/10.1109/ACC.2012.6315687}{\nolinkurl{doi:10.1109/ACC.2012.6315687}}.

\bibitem{ahmadi2013complexity}
A.~A. Ahmadi, A.~Majumdar, and R.~Tedrake, ``Complexity of ten decision
  problems in continuous time dynamical systems,'' in \emph{2013 American
  Control Conference}.\hskip 1em plus 0.5em minus 0.4em\relax IEEE, 2013,
  \href{https://doi.org/10.1109/ACC.2013.6580838}{\nolinkurl{doi:10.1109/ACC.2013.6580838}}.

\bibitem{cai2024dissipative}
Y.~Cai and G.~Pogudin, ``Dissipative quadratizations of polynomial {ODE}
  systems,'' in \emph{Tools and Algorithms for the Construction and Analysis of
  Systems}.\hskip 1em plus 0.5em minus 0.4em\relax Springer, 2024,
  \href{https://doi.org/10.1007/978-3-031-57249-4_16}{\nolinkurl{doi:10.1007/978-3-031-57249-4_16}}.

\bibitem{carr1981applications}
J.~Carr, \emph{Applications of Centre Manifold Theory}.\hskip 1em plus 0.5em
  minus 0.4em\relax Springer-Verlag, 1981,
  \href{https://doi.org/10.1007/978-1-4612-5929-9}{\nolinkurl{doi:10.1007/978-1-4612-5929-9}}.

\bibitem{blondel2000survey}
V.~D. Blondel and J.~N. Tsitsiklis, ``A survey of computational complexity
  results in systems and control,'' 2000,
  \href{https://doi.org/10.1016/S0005-1098(00)00050-9}{\nolinkurl{doi:10.1016/S0005-1098(00)00050-9}}.

\bibitem{braatz1994complexity}
R.~P. Braatz, P.~M. Young, J.~C. Doyle, and M.~Morari, ``Computational
  complexity of {$\mu$} calculation,'' \emph{IEEE Transactions on Automatic
  Control}, vol.~39, no.~5, 1994,
  \href{https://doi.org/10.1109/9.284879}{\nolinkurl{doi:10.1109/9.284879}}.

\bibitem{fu1997hardly}
M.~Fu, ``The real structured singular value is hardly approximable,''
  \emph{IEEE Transactions on Automatic Control}, vol.~42, no.~9, 1997,
  \href{https://doi.org/10.1109/9.623094}{\nolinkurl{doi:10.1109/9.623094}}.

\bibitem{hinrichsen1986structured}
D.~Hinrichsen and A.~J. Pritchard, ``Stability radius for structured
  perturbations and the algebraic {Riccati} equation,'' \emph{Systems \&
  Control Letters}, vol.~8, no.~2, 1986,
  \href{https://doi.org/10.1016/0167-6911(86)90068-X}{\nolinkurl{doi:10.1016/0167-6911(86)90068-X}}.

\bibitem{burgisser2019homology}
P.~B{\"u}rgisser, F.~Cucker, and P.~Lairez, ``Computing the homology of basic
  semialgebraic sets in weak exponential time,'' \emph{Journal of the ACM},
  vol.~66, no.~1, 2019,
  \href{https://doi.org/10.1145/3275242}{\nolinkurl{doi:10.1145/3275242}}.

\bibitem{haddad2008nonlinear}
W.~M. Haddad and V.~Chellaboina, \emph{Nonlinear Dynamical Systems and Control:
  A {Lyapunov}-Based Approach}.\hskip 1em plus 0.5em minus 0.4em\relax
  Princeton University Press, 2008,
  \href{https://doi.org/10.1515/9781400841042}{\nolinkurl{doi:10.1515/9781400841042}}.

\bibitem{schaefer2017fixed}
M.~Schaefer and D.~{\v{S}}tefankovi{\v{c}}, ``Fixed points, {Nash} equilibria,
  and the existential theory of the reals,'' \emph{Theory of Computing
  Systems}, vol.~60, no.~2, 2017,
  \href{https://doi.org/10.1007/s00224-015-9662-0}{\nolinkurl{doi:10.1007/s00224-015-9662-0}}.

\bibitem{abrahamsen2018gallery}
M.~Abrahamsen, A.~Adamaszek, and T.~Miltzow, ``The art gallery problem is
  {$\exists\mathbb{R}$}-complete,'' \emph{Journal of the ACM}, vol.~69, no.~1,
  2022,
  \href{https://doi.org/10.1145/3486220}{\nolinkurl{doi:10.1145/3486220}}.

\bibitem{lasalle1960extensions}
J.~P. LaSalle, ``Some extensions of {Liapunov}'s second method,'' \emph{IRE
  Transactions on Circuit Theory}, vol.~7, no.~4, 1960,
  \href{https://doi.org/10.1109/TCT.1960.1086720}{\nolinkurl{doi:10.1109/TCT.1960.1086720}}.

\bibitem{schaefer1978complexity}
T.~J. Schaefer, ``The complexity of satisfiability problems,'' in
  \emph{Proceedings of the Tenth Annual ACM Symposium on Theory of
  Computing}.\hskip 1em plus 0.5em minus 0.4em\relax ACM, 1978,
  \href{https://doi.org/10.1145/800133.804350}{\nolinkurl{doi:10.1145/800133.804350}}.

\bibitem{hoeven2013bit}
J.~van~der Hoeven and G.~Lecerf, ``On the bit-complexity of sparse polynomial
  and series multiplication,'' \emph{Journal of Symbolic Computation}, vol.~50,
  2013,
  \href{https://doi.org/10.1016/j.jsc.2012.06.004}{\nolinkurl{doi:10.1016/j.jsc.2012.06.004}}.

\bibitem{hillar2013most}
C.~J. Hillar and L.-H. Lim, ``Most tensor problems are {NP}-hard,''
  \emph{Journal of the ACM}, vol.~60, no.~6, 2013,
  \href{https://doi.org/10.1145/2512329}{\nolinkurl{doi:10.1145/2512329}}.

\bibitem{schaefer2018tensor}
M.~Schaefer and D.~{\v{S}}tefankovi{\v{c}}, ``The complexity of tensor rank,''
  \emph{Theory of Computing Systems}, vol.~62, no.~5, 2018,
  \href{https://doi.org/10.1007/s00224-017-9800-y}{\nolinkurl{doi:10.1007/s00224-017-9800-y}}.

\end{thebibliography}
\end{document}